\documentclass[twocolumn,showpacs,prl]{revtex4}

\usepackage{graphicx}%
\usepackage{dcolumn}
\usepackage{amsmath}
\usepackage{latexsym}

\begin {document}

\title
{
Sandpile model on a quenched substrate generated by kinetic self-avoiding trails
}
\author
{
R. Karmakar and S. S. Manna
}
\affiliation
{
Satyendra Nath Bose National Centre for Basic Sciences 
    Block-JD, Sector-III, Salt Lake, Kolkata-700098, India
}

\begin{abstract}
Kinetic self-avoiding trails are introduced and used to generate a substrate
of randomly quenched flow vectors. Sandpile model is studied on such a substrate
with asymmetric toppling matrices where the precise balance between the net outflow of grains from a
toppling site and the total inflow of grains to the same site when all its
neighbors topple once is maintained at all sites. Within numerical accuracy
this model behaves in the same way as the multiscaling BTW model.
\end{abstract}
%\pacs {05.10.-a, 05.40.-a, 05.50.+q, 87.18.Sn}
\maketitle

   Accurate determination of the critical exponents and thereby precise distinction of
critical behaviors among the different universality classes are always regarded as very
important tasks in the field of critical phenomena since this study
helps in understanding as well as identification of the crucial factors that determine
the critical behaviors. This problem however is still open in the phenomenon
of Self-organized criticality (SOC) in spite of extensive research over last several 
years. More precisely, in the sandpile model of SOC the question if the two very important
models namely the deterministic model by Bak, Tang and Wiesenfeld (BTW) \cite {Bak} and
the stochastic Manna sandpile \cite {Manna} belong to the same universality class or not has
not been fully settled yet. A number of works claimed that they belong to the same
universality class \cite {Pietronero, Vespignani, Chessa}, where as a number of other
papers \cite {Ben-Hur, Lubeck, Biham2} argue in favor of their
universality classes being different. However, what was very much lacking till recently is the
precise identification of a key factor which may control the two behaviors.

%  In a recent paper it has been possible to identify such a quantity  \cite {..}.
%It has been argued that a precise toppling balance at every site between the
%number of grains $H$ that are distributed among the neighbours in a toppling
%and the total number of grains $H'$ that are received when all neighbouring sites
%topple for once leads to the BTW universality class where as the absence of this
%precise balance even at a small fraction of sites leads to the Manna universality class \cite {..}.

  In SOC \cite{Bakbook, Dhar1} a system evolves to a critical state by a 
self-organizing dynamics under a constant, slow external drive in the absence of a fine 
tuning parameter. The signature of the critical state is the spontaneous emergence of 
long ranged spatio-temporal correlations in the stationary state. The concept of SOC, 
has been used to explain non-linear transport processes of physical entities like mass, 
energy, stress etc. in phenomena like sandpiles \cite {Bak,Manna}, earthquakes \cite{Sornette}, 
forest fires \cite{Drossel}, biological evolution \cite{Baksneppen} etc. The transport 
manifests itself as intermittent activity  bursts called avalanches. Sandpile models
are the prototype models of SOC.
In spite of extensive efforts BTW model resisted to follow the finite size scaling (FSS) 
ansatz and has been shown recently to obey a multiscaling behavior \cite{Stella1,Stella2}. 
On the other hand scaling behavior in the Manna stochastic sandpile 
\cite{Manna} is observed to be well behaved \cite{Stella2,Lubeck,Biham2}. 
%Whether or not one can put the BTW and the 
%Manna model in the same universality class is an open controversy. 

   A deterministic sandpile model can be defined suitably on an arbitrary
graph by an integer toppling matrix (TM) $\Delta$ \cite {Dhar1}. For example, on a square
lattice of linear size $L$, the number of sand grains at site $i$ is denoted by
$h_i$. Sand grains are added to the system one by one as: $h_i \to h_i +1$.
A threshold value $H_i$ of the number of grains is associated with every site.
A toppling occurs at the site $i$ when $h_i > H_i$. After the toppling the
system is updated using the TM of size $L^2 \times L^2$ as:
$h_j \to h_j - \Delta_{ij}$ for $j=1$ to $L^2$,
where $\Delta_{ii} = H_i >0$ for all $i$ and $\Delta_{ij} \le 0$ for all $i \ne j$.
Therefore during a toppling at the site $i$, the number of grains at this site
is reduced by $\Delta_{ii}$ where as $-\Delta_{ij}$ number of grains flow out to all
sites $j$, $j$ = 1 to $L^2$. A toppling at one
site may make some of its neighboring sites unstable, which may trigger
topplings at further neighborhood, thus creating an avalanche of topplings
in a cascade. The BTW model is a special case of deterministic sandpiles
where the TM has a simple structure like $H_i=4$ and $\Delta_{ij} = \Delta_{ji} = -1$
for each bond $(ij)$, otherwise zero \cite {Dhar1}.       
In the Manna stochastic sandpile model each grain of
the toppling site is transferred to a randomly selected neighboring site implying
that the TM has the annealed randomness and the elements of $\Delta$ matrix 
is constantly updated during the whole course of a given avalanche.

Recently a single sandpile model with quenched random 
toppling matrices is proposed which captures the crucial features of different sandpile models
\cite {Rumani}. In this model the elements of the TM are quenched random
variables, once their values are selected in the beginning, they remain
unchanged. The dynamics of the sandpile is followed with this TM
and the data for avalanches are averaged over different random realizations of 
TMs.  In this model, there can be two possible situations. In the `undirected' case the
TM is symmetric i.e., $\Delta_{ij}=\Delta_{ji}$ where as in the `directed' case
the TM is asymmetric i.e., $\Delta_{ij} \ne \Delta_{ji}$ in general. 
Here, $\Delta_{ij}$ is nonzero only for $i=j$ and for each bond of the lattice.
It is argued that the behavior of undirected model
is similar to the BTW model where as that of the directed model
is similar to the Manna model. 
The distinction between the two models is made even more precise by 
defining two quantities like $H_i=-\sum_{j \neq i} \Delta_{ij}$ i.e., the total number of grains distributed
to the neighbors in a single toppling and
$H_i^{'}=-\sum_{j \neq i} \Delta_{ji}$ which is the number of grains received by the
site $i$ when its every neighbor $j$ topple for once. It has been suggested that
the precise balance
at all sites (except at the boundary sites)
\begin {equation}
H_i = H_i^{'}
\end {equation}
ensures that the model obeys the same multiscaling as in the BTW model. For the directed model this
precise balance is absent in general and the model
shows FSS with the same exponents as in the Manna sandpile model.

%---------------------------------------------------------------------------
\begin{figure}[top]
\begin{center}
\includegraphics[width=6.5cm]{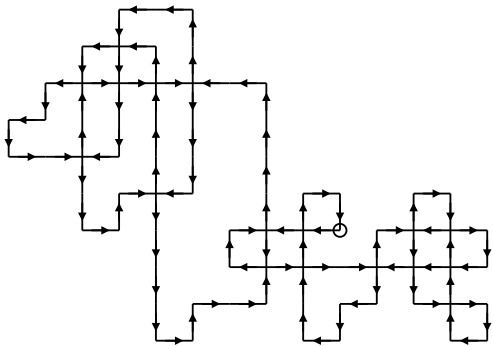}
\end{center}
\caption{ 
A KSAT loop on the square lattice starting from the encircled site and
coming back to the same site after 90 steps.
}
\end{figure}
%---------------------------------------------------------------------------

  In this paper we extend the results of this paper \cite {Rumani} and 
make the conclusion even more precise. We claim that only the precise balance $H_i=H_i^{'}$
or the absence of it determines if a model sandpile would belong to the BTW or Manna
universality class, irrespective of the TM being symmetric or asymmetric.

   A quenched configuration of random flow vectors which corresponds to an
asymmetric TM whose elements satisfy eqn. (1) is generated in the following way. 
Let the neighbors of the site $i$ be denoted by 1, 2, 3 and 4.
We first observe that if one increases the $\Delta$ value of any one of the four 
outgoing bonds, say $(i3)$ by an amount $\delta$, the bond $(i3)$ becomes asymmetric
and it increases $H_i$ by the same amount. Similarly if we increase the $\Delta$ value of an
arbitrary incoming bond to the site $i$, say $(2i)$ by $\delta$ again, the bond $(2i)$ also
becomes asymmetric and $H_i^{'}$ increases by an amount $\delta$. Therefore as a result of both the operations 
the precise balance of $H_i=H_i^{'}$ is strictly maintained.
In general a
series of such bond asymmetrizations can be done randomly by starting from any arbitrary site $i$,
selecting randomly an arbitrary outgoing bond $(ij)$, increasing $\Delta_{ij}$ by $\delta$,
going to the site $j$, selecting an arbitrary outgoing bond $(jk)$ $(\ne (ji))$ and
increasing $\Delta_{jk}$ also by the same amount $\delta$, then going to the site $k$ and so on.
The path obviously cannot visit a bond of the lattice more than once and the
final point to stop must be the starting point. Such a path can intersect
itself but always one of the outgoing bonds which has not been asymmetrized yet is selected
randomly. Since at each site on the path
the $\Delta$ values of either a single or a double pair of incoming and outgoing bonds have been increased
by the same amount $\delta$ the balance of $H_i=H_i^{'}$ is maintained at all sites
on the path.

%---------------------------------------------------------------------------
\begin{figure}[top]
\begin{center}
\includegraphics[width=6.5cm]{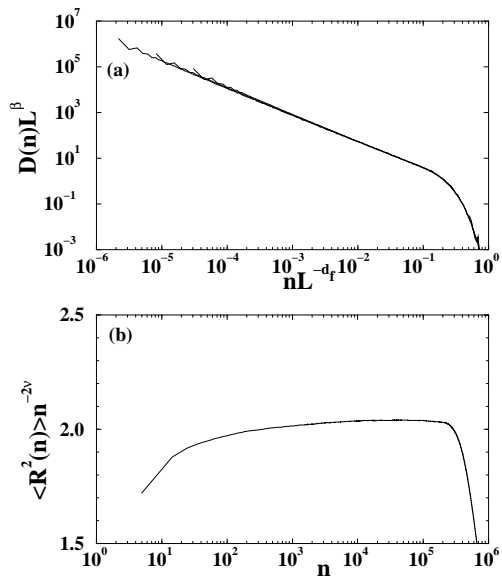}
\end{center}
\caption{ 
(a) Scaling of KSAT loop length distribution for system sizes $L$ = 513, 1025 and 2049.
(b) The mean end-to-end distance of KSATs after $n$ steps grows as $n^{2\nu}$.
}
\end{figure}
%---------------------------------------------------------------------------

   A self-avoiding trail is a random walk which does not visit one bond of the
lattice more than once \cite {Malakis}. A random configuration of self-avoiding trail is
generated by growing a random walk which terminates when a bond is visited more than once.
In contrast a kinetic self-avoiding trail (KSAT) is executed with a little more intelligence.
At each site, to make a step, the walker first finds out the subset of bonds which has not 
been visited yet and then steps randomly along any one of these bonds with equal probability.
Such a walk can also terminate only when it visits the origin for the third time (Fig.1).        
A similar definition of kinetic growth walk or growing self-avoiding walks have been
studied in the literature and it is argued that very long such walks behave in the same
way as ordinary self-avoiding walks \cite {Hemmer}. 

   First we observe that KSATs have very interesting and non-trivial statistics. For example
the probability distribution that a KSAT returns to the origin for the first time after $n$ steps has a
scaling form like:
\begin {equation}
D(n) \sim L^{-\beta}{\cal G}(n/L^{d_f})
\end {equation}
where the scaling function ${\cal G}(x) \sim x^{-\gamma}$ as $x \to 0$ 
such that $\gamma = \beta /d_f$ and ${\cal G}(x) \to$ decreases to zero very fast
when $x \to 1$. We estimated $d_f \approx 1.905$, $\beta \approx 2.237$ which give $\gamma \approx 1.174$ (Fig. 2(a)).
The cut-off exponent $d_f$ is also recognized as the fractal dimension of the KSATs since the number of steps
on the walks whose sizes are of the order of $L$ varies as $L^{d_f}$.
One can also measure
the value of $d_f$ directly. The mean square end-to-end distance $\langle R^2(n) \rangle$ of the walker from the
origin after $n$ steps varies as $n^{2\nu}$, where $\nu = 1/d_f$.
Simulation of walks of lengths up to a million steps on a lattice of size $L$ = 4097 gives
$\nu \approx 0.530$ so that $d_f \approx 1.886$ (Fig. 2(b)). Therefore we conclude a mean value 
of $d_f \approx 1.895$.

   KSATs are therefore used to asymmetrize the TM. We start with a TM whose all elements are
initially zero corresponding to a periodic $L \times L$ lattice. The walker starts from an arbitrarily
selected site, executes a KSAT which finally stops when it comes back to the origin for the first time.
The $\Delta$ values of every outgoing bond visited from each site are then increased by $\delta$
which is selected as a random integer number between 1 and 2. A number of such KSAT loops are then
generated one by one starting from arbitrarily selected sites and with randomly selected $\delta$ values.
The process stops only when all bonds are asymmetrized at least once. The periodic boundary condition
is then lifted. The TM so generated is 
asymmetric in $\approx$ 92.5 \% bonds but maintains the precise balance of $H_i=H_i^{'}$ strictly at all sites except on the
boundary. The lattice is now ready to study the sandpile model where the  threshold height at each site 
is denoted by $H_i$. Such a system has a large fluctuation of threshold heights and their average
increases with increasing the system size.

%---------------------------------------------------------------------------
\begin{figure}[top]
\begin{center}
\includegraphics[width=6.0cm]{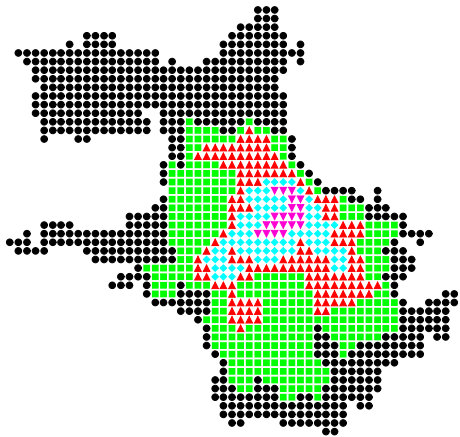}
\end{center}
\caption{(Color online)
Detailed structure of an avalanche, different
sites have toppled different number of times: 1(circle),
2(square), 3(triangle up), 4(diamond) and 5(triangle down).
}
\end{figure}
%---------------------------------------------------------------------------

   We studied three aspects of the sandpile model on the quenched substrate generated by KSATs
which are: (i) the inner structure of the avalanches (ii) the avalanche statistics and the
(iii) wave size distributions. We observe very close similarities of our model with BTW model
in all three aspects as reported below.

   Like any ordinary sandpile model, the dynamics starts from an arbitrary stable distribution of 
sand heights and then grains are added to the system one by one. The system eventually reaches the 
stationary state when the average height per site fluctuates around a mean value but does not grow
any further. The size of an avalanche is measured by the total number of topplings $s$. In Fig. 3
we show the picture of an avalanche which has no holes. Different sites topple different
number of times but the set of sites which toppled same number of times form a connected zone.
The avalanche has an inward hierarchical structure, the $n$-th toppling zone
is completely surrounded by the $(n-1)$-th toppling zone, with the origin situated within the
maximally toppled zone. This is very similar to avalanche structure in the BTW model \cite {Grass}.

   The finite size scaling behavior of the probability distribution Prob($s,L$) 
of avalanche sizes has the following general form:
\begin{equation}
{\rm Prob}(s,L) \sim L^{-\mu} {\cal F}(\frac {s}{L^{D}} ), \quad
\end{equation}
where the scaling function ${\cal F}(x) \sim x^{-\tau}$ in the limit of $x \to 0$ and 
${\cal F}(x)$ approaches zero very fast when $x \to 1$. It is now known that BTW model 
does not follow this FSS form but has a multiscaling behavior \cite {Stella1,Stella2}
where as the Manna model follows this FSS behavior quite accurately \cite{Lubeck}.

%---------------------------------------------------------------------------
\begin{figure}[top]
\begin{center}
\includegraphics[width=6.5cm]{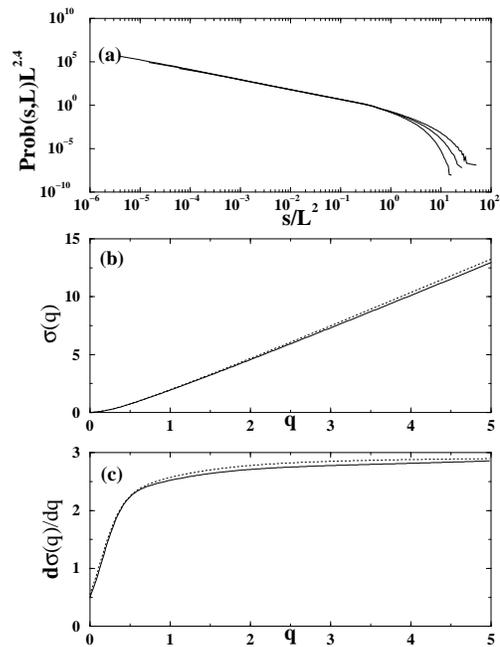}
\end{center}
\caption{
(a) An attempt for the scaling of the  avalanche size distribution
    for the present model for $L$ = 128, 256 and 512.
(b) Comparison of the moment exponents $\sigma (q)$ vs. $q$ and 
(c) $d\sigma (q)/dq$ vs. $q$ for the present model (solid line) and for the BTW (dotted line) model.
}
\end{figure}
%---------------------------------------------------------------------------

For the present model it is observed that the collapse does not work for a single set of $\mu$ and $D$
and for all values of $s$ and $L$. This is a similar situation as found in the 
BTW sandpile model and also in the case of undirected quenched model \cite{Rumani}. For example 
in Fig. 4(a) we have tried an unsuccessful attempt for a data collapse as: {Prob}$(s,L) L^{2.4}$ vs. $sL^{-2}$ 
for $L$ = 128, 256 and 512. Evidently the three curves separate out from one another
beyond $s/L^2 \sim 1$. Even for smaller $s$ values within $1 < s < L^2$ their slopes differ slightly but systematically
as 1.132, 1.135 and 1.144 for $L$ = 128, 256 and 512 respectively, very similar to BTW model behavior.
 
 Further to check that the present model indeed behaves like the multiscaling BTW model
 the various moments of ${\rm Prob}$ are evaluated 
\cite{Stella1,Stella2,Lubeck}.
The $q$-th moment of the avalanche size distribution is defined as 
$\langle s^q \rangle = \Sigma s^q {\rm Prob}(s,L)$.
Assuming that FSS holds for the whole accessible range of avalanche sizes,
it is known that $\langle s^q \rangle \sim L^{\sigma(q)}$
where $\sigma(q) = D(q-\tau+1)$ for $q> \tau -1$
and $\sigma(q)=0$ for $0< q <\tau -1$. 
Estimates of $\sigma(q)$ are obtained from the 
slopes of the plot of $\log \langle s^q (L) \rangle$ with $\log L$ for the three
system sizes mentioned above and for 251 equally spaced $q$ values ranging from 0 to 5.
In Fig. 4(b) we plot $\sigma(q)$ vs. $q$ for the present model and compare it with a
similar plot for the BTW model calculated for the same system sizes, the agreement is found to be very good, within 2\%.
For both models $\sigma(q)$ shows marked deviation from linearity.
To analyze this non-linearity in more detail
it is an usual practice to calculate $d\sigma(q)/dq$ which takes the 
constant value $D$ for large $q$ had the FSS been valid.
In contrast,
in our present case we see in Fig. 4(c) that $d\sigma(q)/dq$
increases steadily with $q$ for $q>1$ and this plot coincides
within the same accuracy with a similar plot for the BTW model.

   In a stable configuration if a grain is
added to a site $i$ with a height $H_i$ it topples and the first
wave is the set of all toppled sites while
site $i$ is prevented from a second toppling. If $i$ is still unstable after
the first wave, the second wave propagates. This process continues until 
site $i$ becomes stable and the avalanche stops \cite{Prie1,Prie2}. 
 
%---------------------------------------------------------------------------
\begin{figure}[top]
\begin{center}
\includegraphics[width=5.5cm]{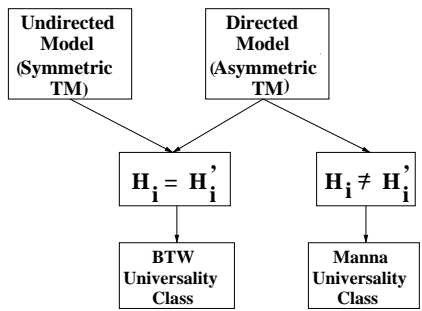}
\end{center}
\caption{ 
This flow chart shows that the precise balance $H_i = H_i^{'}$ or absence of it
determines the universality classes of different sandpile models.
}
\end{figure}
%---------------------------------------------------------------------------

The autocorrelation function of the wave time series $\{s_1,s_2,s_3,\ldots\}$ of
 successive waves \cite{Stella3} is defined as
$C(t,L)=[\langle s_{k+t}s_k \rangle_L-\langle s_k \rangle^2_{L}]/[\langle s^2_k \rangle_L-\langle s_k \rangle^2_L]$ 
 where the $\langle .. \rangle$ refers 
 quenched disorder averaging.
This long range autocorrelation is the consequence 
of the coherent and uniform spatial structure of each wave.  
$C(t,L)$ is found to grow steadily with $L$. It scales as
$C(t,L)\sim t^{-\tau_c} G(t/L^{D_c})$ with same exponents as undirected model 
with $\tau_c \approx 0.35$ and 
$D_c \approx 1$. These exponents should be compared to 0.40 and 1.02, 
respectively, as determined for the BTW model \cite {Stella3}.

   To summarize, an asymmetric toppling matrix is generated using random
kinetic self-avoiding trail loops on the square lattice. The TMs generated
in this way guarantees the precise balance between the outflow of grains during a single toppling
at a site and the total number of grains flowing into the same site when all its
neighbors topple for once. A deterministic sandpile model is studied on such
a quenched random lattice and the statistical behavior of its avalanches 
are compared with that of BTW model in a number of ways, namely,
the inner structure of the avalanches, the avalanche statistics and the
wave size distributions. Within numerical accuracy excellent agreement
is observed in all three categories. We conclude, as displayed in a flow
chart in Fig. 5, that it is only the local
flow balance or absence of it, irrespective of it being generated from a symmetric or
asymmetric TM, determines the universality class of the sandpile model.

\end{document}